\documentclass[12pt]{iopart}
\usepackage[pdftex]{graphicx}
\usepackage{color}
\usepackage{subfigure}
\newcommand{\braket}[2]{\langle{#1} | {#2} \rangle}
\newcommand{\bra}[1]{\left\langle #1 \right|}
\newcommand{\ket}[1]{\left| #1 \right\rangle}

\begin{document}

\setcounter{tocdepth}{2}

\title{Experimental creation and analysis of displaced number states}

\author{F.~Ziesel, T.~Ruster, A.~Walther, H.~Kaufmann, K.~Singer, F.~Schmidt-Kaler and U.~G.~Poschinger}

\address{QUANTUM, Institut f\"ur Physik, Universit\"at Mainz, Staudingerweg 7, 55128 Mainz, Germany}
\ead{poschin@uni-mainz.de}
\begin{abstract}
We create displaced number states, which are non-classical generalizations of coherent states, of a vibrational mode of a single trapped ion. The creation of these states is accomplished by a combination of optical and electrical manipulation of the ion. A number state is first prepared by laser-driven climbing of the Jaynes-Cummings ladder, followed by displacement created by a sudden shift of the electrostatic trapping potential. Number states $n=$0,1 and 2 are prepared, and displacement amplitudes of up to $\alpha\approx 2.8$ are reached. The states are analyzed tomographically, and we find a good agreement with the theoretically expected results for displaced number states. Quantum features elucidating the concept of interference in phase space are clearly demonstrated experimentally.
\end{abstract}

\maketitle
\tableofcontents

\section{Introduction}
In the last decades, it has become possible to perform experiments on well isolated simple quantum systems. For example, a trapped single ion can be manipulated with laser interactions such that non-classical states of the ion's internal electronic and external vibrational states can be created and probed. Such experiments allow for exploring the fascinating features of quantum entanglement \cite{Monroe96,Deleglise08}, and have set the basis for emerging quantum technologies, e.g. scalable quantum processors using ions in a segmented ions trap device \cite{Kielpinski02}. Engineering and probing non-classical states of quantum systems, besides being of interest in its own right, has many prospects and applications. Techniques for quantum control and measurement can be established and benchmarked, decoherence properties of different quantum states can be compared \cite{Barreiro10}, and theoretical concepts such as negativity of quasi probability distributions can be studied. Furthermore, nonclassical states set the ground for quantum simulation experiments \cite{Leibfried02,Gerritsma10} and offer the possibility of enhanced interferometric resolution \cite{Caves81}, providing the basis for future quantum sensors.\\
An example of a nonclassical state is the \textit{displaced number state} (DNS), which is a generalization of the well-known Glauber coherent state. DNS can be represented as $\ket{\alpha,n}$ in Dirac notation, where $n$ is the quantum number of an energy eigenstate of a harmonic oscillator, which is additionally displaced by an amplitude $\alpha$ in phase space. DNS have been studied theoretically \cite{Boiteux73,Oliveria90,Wunsche91,Moya95}, they provide a versatile tool e.g. for the calculation of quasi-probability distributions, as their coherent or number state nature can be exploited depending on the context. Furthermore, DNS provide a clear illustration of the concept of \textit{interference in phase space} \cite{Dowling91}, e.g. yielding a semiclassical explanation of molecular spectra. Experimental work on DNS done so far with photons includes generation and characterization of the state $\ket{\alpha, 1}$ with displacements of up to $\alpha\approx 2.4$ \cite{Lvovsky02}, which revealed clear deviations from the ordinary coherent state $\ket{\alpha,0}$, and signatures of interference in phase space have been observed. A more recent experiment with photons showed the non-classical decrease of the overlap integral $|\braket{ 1}{\alpha,1} |^2$ with increasing $\alpha$ \cite{Laiho10}. Further experimental work established an analogy between DNS and classical light propagation in Glauber-Fock waveguides \cite{Perez10,Keil11}.\\
For trapped ions, many fascinating experiments have been performed on the generation and analysis of nonclassical states in the last two decades, prominent examples being the creation of Fock, coherent and squeezed states \cite{Meekhof96}, the tomographic reconstruction of quasiprobability distributions \cite{Leibfried96} and more recently the entanglement of the motional modes of two trapped ions \cite{Jost09}. Our work is extending the previous work on nonclassical states of trapped ions by adding DNS to the collection of states generated and analyzed so far. While photonic experiments have been limited to states $\ket{\alpha,1}$, we prepare states of up to $n=2$, and we are able to see characteristic higher-order interference minima in phase space. In addition, we demonstrate that both optical and non-resonant electrical forces can be applied together with sufficient precision to allow for quantum state engineering of the motional degree of freedom. We cannot perform a direct phonon number measurement, compared to the direct photon number detection in Ref. \cite{Laiho10}. Thus we rather reconstruct the generated state of the motional mode of a single trapped ion by a tomographic method, which features a high precision and allows for a rather precise matching of the obtained results to the theoretical prediction.

\section{Experimental scheme}

\subsection{Preparation of displaced Fock states}

\begin{figure}[h!]
	\centering
	\qquad\qquad\subfigure{\includegraphics[width=0.36\textwidth]{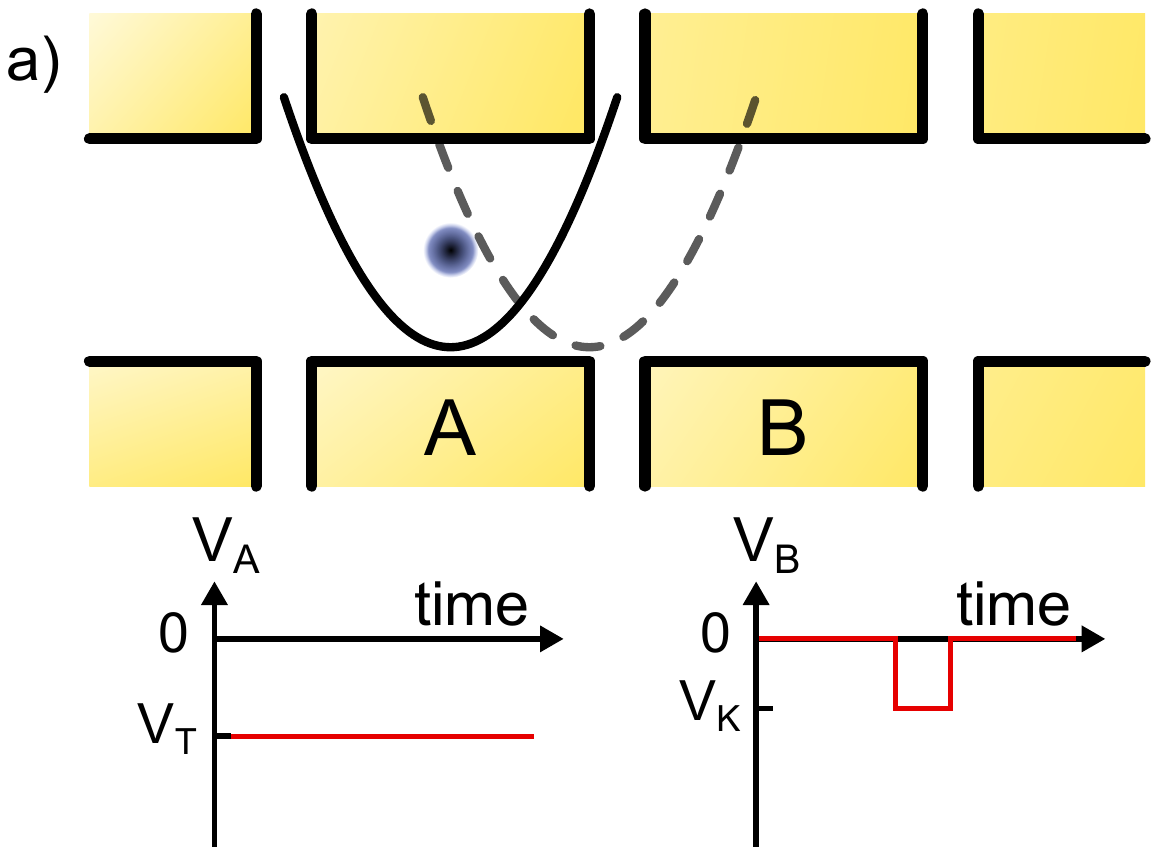}}\hfill
	\subfigure{\includegraphics[width=0.36\textwidth]{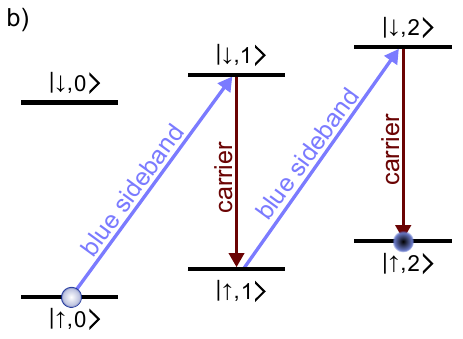}}\qquad\qquad
\caption{State preparation scheme: \textbf{a)} shows how controlled displacement is achieved by switching the voltage of the trap segments. The ion is initially kept at segment A, while a voltage pulse on segment B temporally shifts the potential minimum. \textbf{b)} illustrates the preparation of number states by alternating $\pi$-pulses on blue sideband and carrier transitions.}
\label{fig:sketch}
\end{figure}

To create DNS of the axial mode of vibration of a single trapped ion, we combine different methods for the quantum control of motional and internal degrees of freedom. We first create number states by a sequence of laser pulses consisting of $\pi$ pulses on the blue motional sideband, each adding a phonon to the motional excitation, and $\pi$ pulses on the carrier transition to reset the internal state. Electrical forces generated with control voltages at the trap segments are then used to displace the harmonic oscillator by a well defined amplitude. The scheme is sketched in Fig. \ref{fig:sketch}. The displacement operation is accomplished by the application of a voltage kick generated by a fast arbitrary waveform generator. Such hardware technology with sufficient precision and low noise level became available only recently, enabling the fast transport of trapped ion over long distances \cite{Bowler12,Walther12}. We note that displacement can also be accomplished by different means. One possibility is the application of optical forces \cite{Monroe96}, here the magnitude of achievable displacement is limited by the requirement that the oscillation amplitude in position space has to be well below the wavelength of the optical driving field \cite{Poschinger10}. Another possibility is resonant electrical driving \cite{Leibfried96}.\\
The detailed experimental scheme is as follows: A single $^{40}$Ca${^+}$ ion is stored in a microstructured segmented Paul trap \cite{Schulz08} and laser cooled on the $S_{1/2}\leftrightarrow P_{1/2}$ cycling transition near 397~nm. The $m_J=+1/2$ and $m_J=-1/2$ sublevels of the $S_{1/2}$ ground state are henceforth termed $\ket{\uparrow}$ and 
$\ket{\downarrow}$, respectively. These levels are split in energy by an external magnetic field, leading to a Zeeman splitting of about $\omega_Z\approx 2\pi\cdot$18~MHz. Coherent transitions between these states are driven via stimulated Raman transitions \cite{Poschinger09}. These transitions are mediated via the $P_{1/2}$ state and driven by two optical fields near 397~nm, red detuned from the cycling transition by a frequency of about $\Delta_R\approx 2\pi\cdot$40~GHz. Coupling to the axial mode of vibration takes place with a Lamb-Dicke factor of about $\eta\approx 0.21$ \cite{Leibfried03}, allowing for the coherent driving of motional sidebands.\\
For each measurement, the axial mode of vibration at a frequency of $\omega_{\mathrm{ax}}\approx 2\pi\cdot$1.35~MHz is cooled close to the motional ground state with pulsed Raman sideband cooling on a stimulated Raman transition, reaching a mean phonon number as low as $\bar{n}\leq 0.1$. The internal state is then initialized to $\ket{\uparrow}$ by optical pumping. We then prepare a Fock state $\ket{n}$ by performing a $\pi$-pulse on the blue motional sideband of the stimulated Raman transition, changing the state according to 
\begin{equation}
\ket{\uparrow,m}\rightarrow\ket{\downarrow,m+1}.
\end{equation}
The internal state is then reset by a $\pi$-pulse on the carrier transition, 
\begin{center}
\begin{equation}
\ket{\downarrow,m+1}\rightarrow\ket{\uparrow,m+1}. 
\end{equation}
\end{center}
This sequence is repeated $n$ times. For each Fock state preparation step $m$, the pulse areas for the sideband and carrier pulses are calibrated by monitoring the respective Rabi oscillations. The preparation is concluded by additional optical pumping to $\ket{\uparrow}$ with a circularly polarized beam at the cycling transition. The fidelity of the Fock state preparation is limited by the imperfect preparation of the motional ground state, leading to a fidelity deterioration for higher $n$ , since the transfer pulse areas depend on $m$ and are therefore not well defined. This results in a spread of the population over the state ladder. The concluding optical pumping facilitates the evaluation of the data by setting a well defined internal state of the ion, which compensates for dispersion-induced errors of pulse areas of the preparation pulses. However, deviations of the desired final motional quantum number are not compensated, this error is even slightly increased as there is a finite probability of changing the motional state in an excitation and decay cycle of the pumping.\\
The displacement is then performed by changing the voltage levels on the trap segment and its neighboring segment, further details are explained in Sec. \ref{sec:displacement}. The voltages are switched back to the initial values after 400~ns. We then apply an analysis pulse of variable duration on the stimulated Raman transition and its respective blue and red sidebands. The analysis procedure is further detailed in Sec. \ref{sec:analysis}. Finally, the population in $\ket{\uparrow}$ is selectively transferred to the metastable $D_{5/2}$ state by electron shelving with laser light near 729~nm. Upon subsequent saturated driving of the cycling transition, fluorescence will be observed if the ion is found to be in the $\ket{\downarrow}$ state before the shelving, and no fluorescence will be observed otherwise, such that a spin-selective readout is performed. The entire sequence is repeated 200 times to obtain an estimate of the probability $P_{\uparrow}$ to be in the $\ket{\uparrow}$ state. This measurement is carried out for various analysis pulse durations and analysis pulse detunings, corresponding to resonant driving of red sideband, carrier and blue sideband. 

\subsection{State analysis}
\label{sec:analysis}
We describe the motional state of the ion, which is not necessarily a pure state, by the density matrix $\hat{\rho}$. We assume preparation of the internal state $\ket{\uparrow}$, thus the total state is given by $\ket{\uparrow}\otimes \hat{\rho}$ prior to the analysis pulse. The final probability to find the ion in the $\ket{\uparrow}$ state is then given by the Rabi oscillation signal
\begin{equation}
P_{\uparrow,\Delta n}(\theta )=\frac{1}{2}\sum_{k=0}^{k_{\mathrm{max}}}p_k\left(1+\cos(M_{k,\Delta n}\theta)\right),
\end{equation}
where $\theta$ indicates the analysis pulse area and $p_k=\mathrm{Tr}\left(\hat{\rho}\ket{k}\bra{k}\right)$ is the phonon probability distribution (PPD). This is an incoherent superposition of Rabi oscillations starting from different number states $k$, where the Rabi frequency is altered by the matrix element $M_{k,\Delta n}$~\cite{Leibfried03} in a nonlinear way. This opens up the possibility to infer $p_k$ from the measured signal. The reconstruction of the PPD from a set of measured functions, $P_{\uparrow}(\theta)$, with $\delta n=$~-1,0,1, is done with a maximum likelihood parameter estimation, where the $p_k$ values for $k\leq k_{\mathrm{max}}=6$ are varied freely, besides the requirement of a normalized PPD. Further parameters such as the bare Rabi frequency and the readout fidelity are included in the reconstruction to suppress systematic errors. 

\subsection{Controlled displacement by fast voltage switching}
\label{sec:displacement}


The displacement operation $\ket{n}\rightarrow \hat{D}(\alpha)\ket{n}=\ket{\alpha,n}$ is performed by switching the voltage on the  neighboring segment to the trap segment, with its center located 280~$\mu$m away from the initial trap position. A voltage of 1.0~V at this segment will give rise to an electric field of 600~V/m at the initial ion location, which is derived from electrostatic simulation of the trap \cite{Singer10}. This field exerts an acceleration of approximately 1.5$\cdot$10$^9$~m/s$^2$. Assuming an ideal, sudden switching process, the potential minimum is shifted by a well defined distance along the trap axis for well defined time.  The relation between the set voltage and the magnitude of the shift of the potential minimum is established via electrostatic simulation of the trap and has been experimentally confirmed \cite{Huber10}. 
For the displacement, the set voltage is changed instantaneously, however $\Pi$-type filters with cutoff frequencies of approximately 300~kHz lead to smoothened voltage ramps at the trap segment. The actual displacement amplitude after the voltage kick is therefore lower than the value corresponding to the position space shift of the potential minimum, furthermore the displacement pulse is delayed and blurred. With the deviation of the potential minimum from its original location $x_0(t)$, the resulting displacement is obtained from classical harmonic oscillator dynamics\cite{Bowler12}:
\begin{equation}
\alpha(t)=-\sqrt{\frac{m\omega}{2\hbar}}\left(e^{-i\omega t}\int_0^{t} \dot{x}_0(\tau)e^{i\omega\tau}d\tau\right).
\end{equation}



The fast switching of the voltages is done with a homebuilt fast multichannel arbitrary waveform generator. Its main building blocks are quad 16-bit serial digital to analog converters \footnote{DAC8814, Texas Instruments}, which cover the output voltage range of -10~V to +10~V with a resolution of about 0.3~mV. The digital data is supplied in real time by a field programmable gate array (FPGA) \footnote{Avnet Virtex-5 FXT Evaluation Kit}, which in turn is programmed via a Gigabit Ethernet connection from a control computer. The maximum update rate is limited by the clock rate of the FPGA and is given by 2.5~MSamples/s, and a number of up to 48 output channels can be controlled individually. A crucial feature is the possibility to vary the duration of each individual sample in steps of 20~ns beyond the minimum value of 400~ns, which allows for resolving the trap oscillation period \cite{Walther12}.\\
Each output channel at voltage $V_i$ is fed into a stage where a differential voltage $V_d$ is added (subtracted) such that two channels with output voltages $V_i\pm V_d$ are obtained, which are fed to adjacent segments at the same axial location in the trap, which allow for the compensation of micromotion. The differential voltage $V_d$ can be derived from a DAC, or from an external input.\\
The measured output voltage noise at the trap frequency, in the steady state, is as low as 1.0$\cdot$10$^{-9}$~V/$\sqrt{Hz}$. This results in a spectral density of the electric field noise of about 3.6$\cdot$10$^{-13}$~V$^2$~/~Hz~m$^2$, which in turn yields a heating rate that lower than the actually observed heating rate of $\dot{n}\approx 0.3$~ms$^{-1}$\cite{Poschinger09} by one order of magnitude; it therefore does not significantly contribute to the anomalous heating. Care was take to suppress digital noise, i.e. feedthrough of the FPGA clock and other digital signals onto the output channels is reduced by using high speed magnetic insulators for all digital signal lines and careful separation of analog and digital signals on the circuit boards bearing the DACs.\\
We predict the resulting displacement magnitudes from simulations, which take into account the actual voltage waveforms applied to the shift segment. These waveforms are measured behind the filter. Averaging of about 10$^5$ identically generated waveforms was necessary to make the results independent of measurement artifacts. These waveforms are used in conjunction with the simulated electrostatic potentials for the numerical solution of the classical equation of motion,
\begin{equation}
m\ddot{x}=-e~U_A~\frac{dV_A^{(1V)}}{dx}-e~U_B(t)~\frac{dV_B^{(1V)}}{dx},
\end{equation}
where the forces are given by the spatial derivatives of the normalized electrostatic potentials $V_i^{(1V)}(x)$ along the trap axis, which result from a segment voltage of 1~V applied to electrode $i$. These forces are weighted by the respective voltages applied to the electrodes and added. The final energy $E_f$ at times larger than the settling time is rescaled to harmonic oscillator units, $E_f'=E_f/\hbar\omega$, and the displacement is obtained from $\left|\alpha\right|=\sqrt{E_f'}$. The resulting displacement amplitudes versus the kick amplitude $V_k$ are shown in Fig. \ref{fig:alphaVsV}.

\section{Results}
\begin{figure}[h!]
\centering
\resizebox{\textwidth}{!}{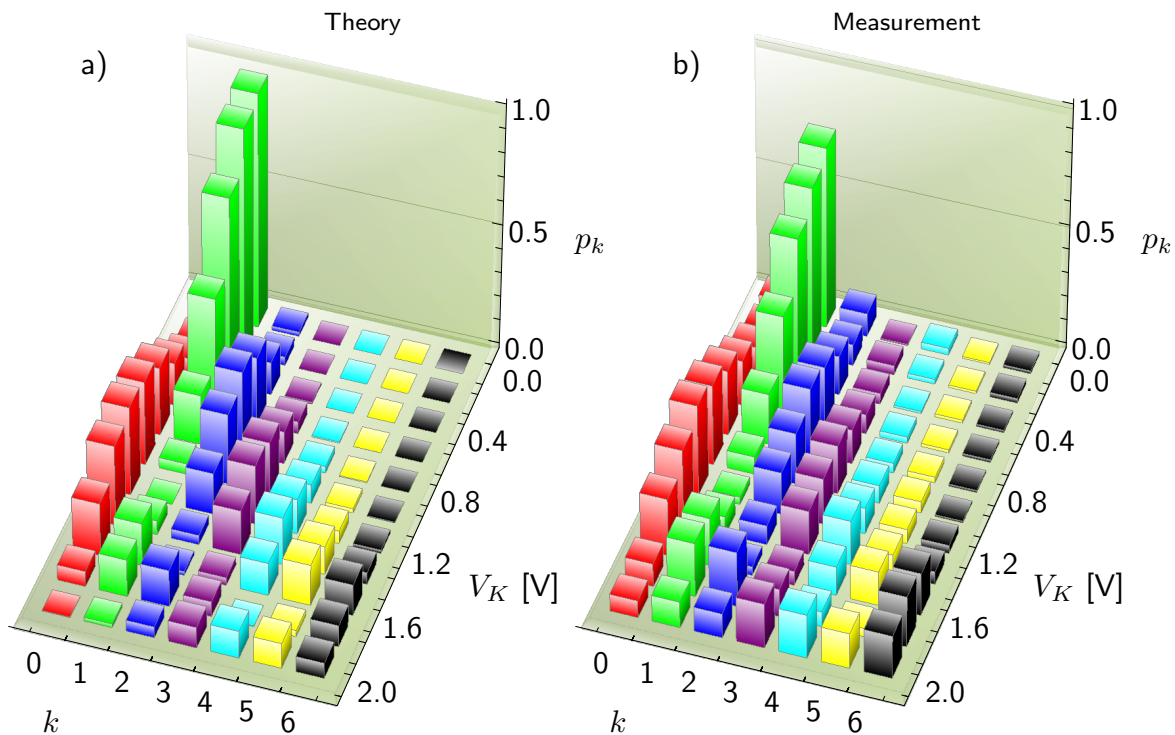}
\caption{Resulting theoretical PPD \textbf{a)} and PPD reconstructed from the measurement result \textbf{b)} for the preparation of $n=1$. The PPD are shown for each applied kick voltage. The interference minimum is clearly observable as a ditch bending toward higher phonon numbers for larger kick voltages.}
\label{fig:coFo1}
\end{figure}

\begin{figure}[h!]\begin{center}
\includegraphics[width=0.7\textwidth]{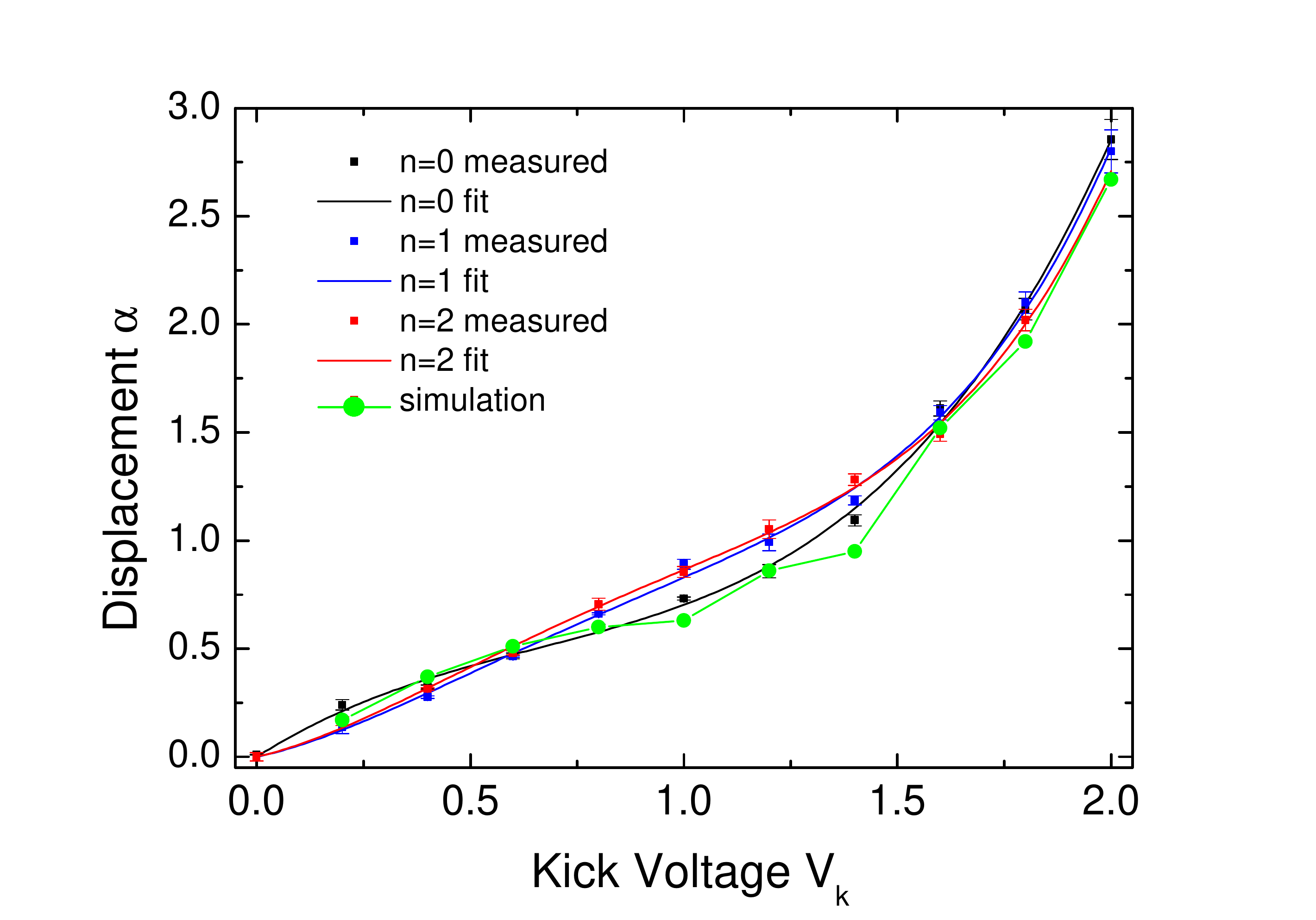}
\caption{Extracted values for the displacement parameter $\alpha$ versus the kick voltage $V_k$ for the datasets pertaining to $n=0$ (black), $n=1$ (blue) and $n=2$ (red). The solid lines are fits to polynomials of fourth order. The displacement parameters are obtained by fitting the theoretical predicition Eq. \ref{eq:ppdconv} to the reconstructed PPDs. The curves are clearly consistent, providing justification for the evaluation method. A slight deviation is observed for the $n=0$ case, which we attribute to a slow drift of the trap voltages. The green points show results from simulations, see text.}
\label{fig:alphaVsV}
\end{center}\end{figure}

\begin{figure}[h!]\begin{center}
\includegraphics[width=\textwidth]{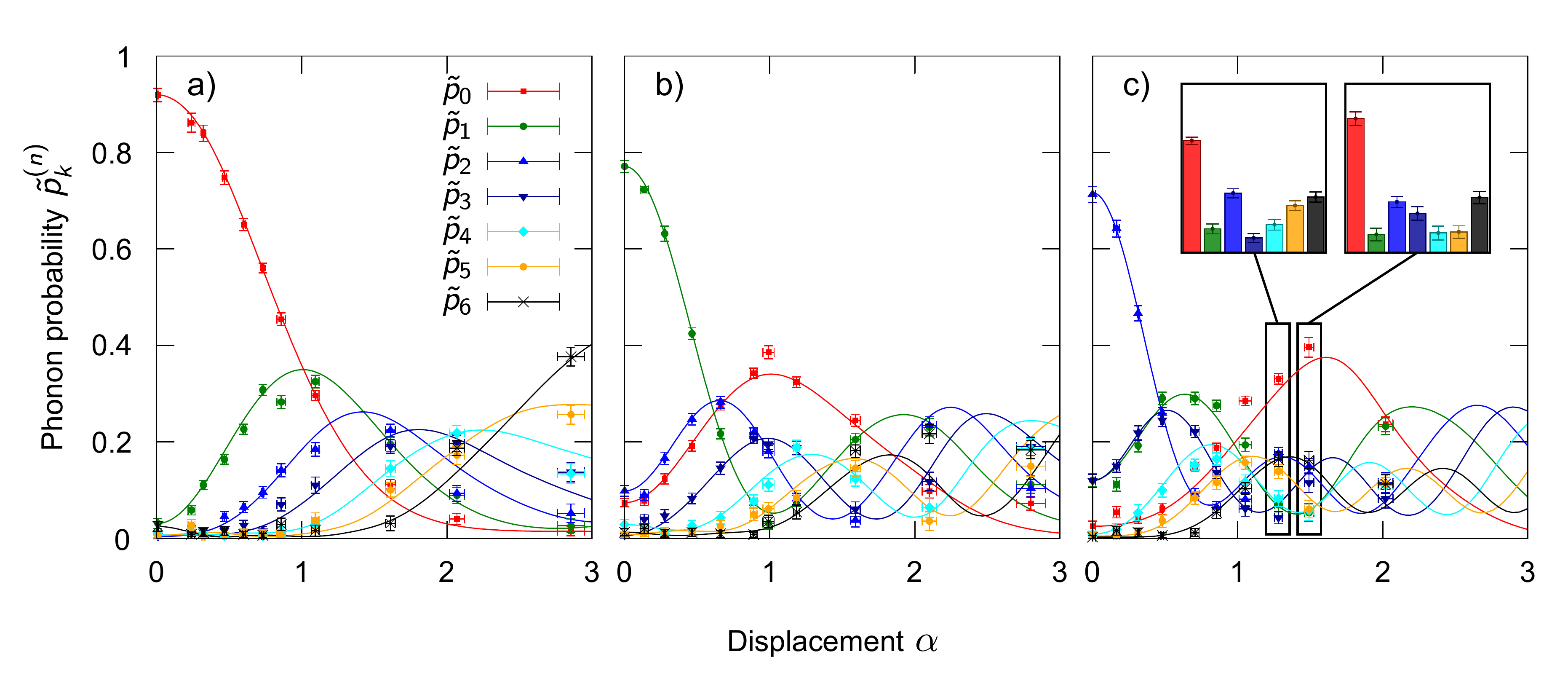}
\caption{Resulting PPDs for different number state preparations \textbf{a)} $n=0$, \textbf{b)} $n=1$, \textbf{c)} $n=2$. The respective PPDs are plotted versus the displacement amplitude $\alpha$, which in turn is extracted by fitting the PPD to Eq. \ref{eq:ppdtheo}. The solid lines display the theoretically expected values from the same Eq. \ref{eq:ppdtheo}. The matching of the experimental values to the theoretical values becomes worse for larger displacement amplitudes because the precision of the maximum likelihood reconstruction is deteriorated by truncation errors. The insets for the $n=2$ show bar charts of the PPDs at $\alpha\approx 1.5$ and $\alpha\approx 1.5$ to clearly show the interference minima.}
\label{fig:coFoAll}
\end{center}\end{figure}

With the optically driven ladder climbing method described above, we prepare number states for $n=0,1,2$ before applying the kick. The displacement is determined by the magnitude of the voltage kick on the neighboring segment. The kick voltage $V_k$ is changed in steps of 200~mV, reaching a maximum displacement of $\alpha\approx 2.1$ at $V_k = 2$ V. In each case, the state reconstruction is performed with freely varying PPD. As a result, we obtain a normalized PPD for each pair of $n$ and $\alpha$. The results for $n=1$ are shown in Fig. \ref{fig:coFo1}. The measured PPD is compared to the theoretically expected one \cite{Boiteux73,Oliveria90},
\begin{equation}
p_k^{(n)} = \left|\braket{k}{\alpha,n}\right|^2 = e^{-|\alpha|^2 } |\alpha|^{2(k-n))}n! k! \times \left| \sum_{l=0}^n \frac{(-1)^{l} |\alpha|^{2(n-l)}}{l! (n-l)! (k-l)!}\right|^2.
\label{eq:ppdtheo}
\end{equation}

\begin{figure}[h!]\begin{center}
\includegraphics[trim=2cm 5cm 6cm 2.5cm, clip=true, width=0.7\textwidth]{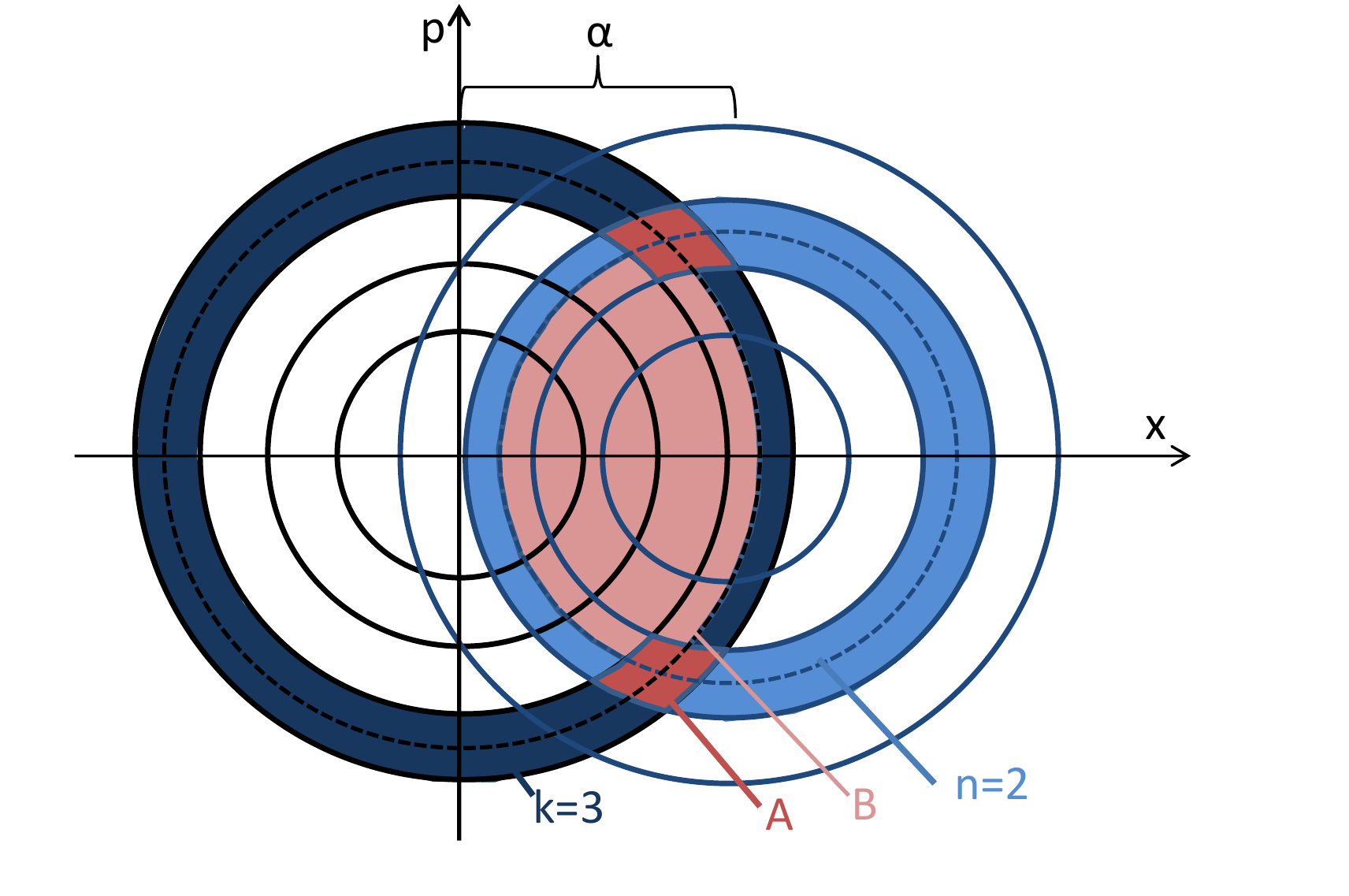}
\caption{Illustration how interference in phase space leads to the observed minima in the PPDs: We exemplify the overlap integral $\braket{3}{\alpha,2}$. The Fock states are represented semiclassically as bands in phase space, spread around the classical trajectories corresponding to the energy eigenvalues $(n+1/2)\hbar\omega$ (dashed lines). The analysis number state basis is represented by bands centered around the origin, the DNS states $\ket{\alpha,n}$ are displaced by the amplitude $\alpha$. The overlap integral is determined not only by the diamond shaped red areas $A$, but also by the intersection area $B$ of the trajectories (shaded in light red), see text.}
\label{fig:interference}
\end{center}\end{figure}

We take into the imperfect preparation of the desired number state prior to the displacement operation. We assume that the action of the displacement kick is not dependent on the quantum state of the motional mode, which is justified by the spatial homogeneity of the electrical force. If we characterize the motional mode after preparation of the desired Fock state $n$ by the density matrix $\hat{\rho}_0^{(n)}$, we obtain the PPD which is actually measured from a convolution of the pure PPD from Eq. \ref{eq:ppdtheo} with the the diagonal elements of the density matrix corresponding to the measurement with zero kick voltage, i.e. the PPD without displacement operation:
\begin{equation}
\tilde{p}_k^{(n)} = \sum_{m=0}^{n_{\mathrm{max}}}  \mathrm{Tr}\left(\hat{\rho}_0^{(n)}\ket{m}\bra{m}\right) p_k^{(m)}.
\label{eq:ppdconv}
\end{equation}
From the reconstructed density matrices without kick, $\hat{\rho}_0^{(n)}$, we determine the respective Fock state preparation fidelities $\mathrm{Tr}\left(\hat{\rho}_0^{(0)}\ket{0}\bra{0}\right)\approx$~0.92, $\mathrm{Tr}\left(\hat{\rho}_0^{(1)}\ket{1}\bra{1}\right)\approx$~0.77, and $\mathrm{Tr}\left(\hat{\rho}_0^{(1)}\ket{1}\bra{1}\right)\approx$~0.72. To prove the consistency of this evaluation method, we show the extracted displacement magnitudes $\alpha$ versus the kick voltage $V_k$ for the three different data sets in Fig. \ref{fig:alphaVsV}. When these values are compared to displacement values inferred from simulations, we find an agreement on the single quantum level.\\
In Fig. \ref{fig:coFo1}, we show the resulting PPDs for the case $n=1$ plotted versus the kick voltage. The results clearly exhibit a single minimum in the PPD, which is moving to higher phonon numbers for larger kick voltages. The occurrence of two minima in the PPD is visible for $n=2$ (see Fig.~\ref{fig:coFoAll}), which has not been observed before in photonic realizations of DNS. The fact that $n$ minima can occur in the PPD of a state $\ket{\alpha,n}$ can be directly read off from Eq. \ref{eq:ppdtheo}: Within the sum a polynomial of order $n$ in $|\alpha|^{2}$ with coefficients of alternating signs is formed, which maximally exhibits $n$ zeros upon squaring \cite{Oliveria90}.\\
To provide a semiclassical explanation for the occurrence of the minima, we illustrate the idea of interference in phase space in Fig. \ref{fig:interference} \cite{Schleich01}. The overlap integral $\braket{k}{\alpha,n}$ is given by the coherent sum of the two intersection areas $A$ between the bands representing the number states $\ket{k}$ and $\ket{n}$. In the summation, one has to take into account the relative phase $\phi_{nk}(\alpha)=B/\hbar$  which is given by the area $B$ enclosed between the two classical trajectories:
\begin{equation}
 \left|\braket{k}{\alpha,n} \right|^2=\frac{2}{\pi\hbar}\cos{\phi_{nk}{\alpha}}.
\end{equation}
As the preparation number $n$ increases, the values of $\phi_{nk}(\alpha)$ sweep over larger ranges for increasing analysis quantum numbers $k$ and increasing $\alpha$, thus allowing for a larger number of interference minima.

\section{Outlook and conclusion}
We have demonstrated the creation of DNS $\ket{\alpha,n}$ with number states of up to $n=2$ and $\alpha\approx$~2.9, going significantly beyond the current state of the art of photonic experiments done on DNS so far. Emphasis is put on the fact that the displacement is created by electrical kicks, not by resonant excitation by rf pulses. This offers the flexibility to combine long-distance shuttling operations and controlled motional excitation on the  single quantum accuracy level, which might open up the route to scalable quantum simulation schemes. The displacement operations were shown to be very accurate and reproducible. We emphasize that we were able to match the measured displacement amplitudes to simulation results on the single quantum accuracy level, where technological details such as the electrostatic properties of the microtrap and the voltage waveforms have to be determined with a sufficient precision.\\
Future applications in the field of quantum state engineering include the joint usage of spin dependent optical forces and electrical forces, e.g. for optically assisted splitting of ion crystals or for phase referencing optical beat patterns. The latter might open up the possibility for novel measurement schemes where an optical interference pattern is to be applied with a fixed phase relation to ion strings \cite{Walther12a}, which could be of use for quantum simulation experiments \cite{Pruttivarasin11}.

\section{Acknowledgements}
The authors acknowledge gratitude to Heinz Lenk for development of electronics hardware. We acknowledge financial support by the IARPA SQIP project (MQCO framework), and by the European commission within the IP AQUTE and STREP Diamant, and the VW-Stiftung. A.W. acknowledges funding from the Swedish Research Council.

\section*{References}
\bibliographystyle{iopart-num}
\bibliography{displacedNumberStates}

\end{document}